# Superconducting phase diagrams of $LuB_{12}$ and $Lu_{1-x}Zr_xB_{12}$ ($x \leq 0.45$) down to 50 mK

J. BACKAI[1], S. GABANI[2,*], K. FLACHBART[2], E. GAZO[2], J. KUSNIR[3], M. ORENDAC[2,3], G. PRISTAS[2], N. SLUCHANKO[4,5], A. DUKHNENKO[6], V. FILIPOV[6], N. SHITSEVALOVA[6]

[1] Faculty of Electrical Engineering and Informatics, Technical University, Letná str. 9, 04200 Košice, Slovakia
[2] Institute of Experimental Physics, Slovak Academy of Sciences, Watsonova str. 47, 04001 Košice, Slovakia
[3] Institute of Physics, Faculty of Science, P. J. Safarik University, Park Angelium str.9, 04154 Košice, Slovakia
[4] Prokhorov General Physics Institute, Russian Academy of Sciences, Vavilov str. 38, 119991 Moscow, Russia
[5] Moscow Institute of Physics and Technology, Moscow Region, 141700 Russia
[6] Institute for Problems of Materials Science, National Academy of Sciences of Ukraine, Krzhyzhanovsky str. 3, 03142 Kiev, Ukraine

Lutetium dodecaboride $LuB_{12}$ is a simple weak-coupling BCS superconductor with critical temperature $T_c \approx 0.42$ K, whilst $ZrB_{12}$ is a strong-coupling BCS superconductor with the highest critical temperature $T_c \approx 6.0$ K among this group of materials. In case of lutetium substitution by zirconium ions in $LuB_{12}$ the crossover from weak- to strong-coupling superconductor can be studied. We have investigated the evolution of critical temperature $T_c$ and critical field $H_c$ in high-quality single crystalline superconducting samples of $Lu_{1-x}Zr_xB_{12}$ ($0 \leq x \leq 0.45$) by measuring magnetic *ac*-susceptibility between $\approx 1$ K and 50 mK. To obtain this kind of experimental data, a new susceptometer was designed, constructed and tested, which can work in a wide temperature range of 0.05 K – 3 K in $^3$He-$^4$He dilution refrigerator. The measurements with this new susceptometer revealed how $T_c(x)$ and $H_c(x)$ increases with increasing concentration of zirconium in $Lu_{1-x}Zr_xB_{12}$ solid solutions as well as how their superconducting phase diagram develops.

## 1. Introduction

Superconductivity in dodecaborides $ZrB_{12}$ ($T_c \approx 6$ K [1]) and $LuB_{12}$ ($T_c \approx 0.4$ K [2]) is well known for more than 40 years. $ZrB_{12}$ and $LuB_{12}$ are very similar in their conduction band, magnetic and crystalline structure. Despite this, it is not clear up to now why there is such a big difference in their critical temperatures and fields. Studies of inelastic neutron scattering of the phonon spectra in $ZrB_{12}$ and $LuB_{12}$ [3] have shown only small changes in the position of their almost dispersion-less quasi-local modes, which is referred to vibrations of loosely bounded Zr or Lu atoms in cubooctahedron cavities of the rigid covalent boron sublattice. Regarding to electron structure of $ZrB_{12}$ and $LuB_{12}$, a moderate increase (about 0.3 - 0.4 eV of the Fermi level [4]) in the electron density of states of zirconium dodecaboride was observed due to two times increase in conduction band filling when Zr ions were replaced by Lu ions.

Detailed investigations of superconducting properties of solid solutions $Lu_{1-x}Zr_xB_{12}$ can shed more light on this problem. As the first step of this study we show how the concentration dependence of critical temperature $T_c(x)$ and temperature dependences of the critical magnetic field $H_c(T)$ for various concentration of zirconium develop. Recently, a rather complex experimental research of electrical resistivity, magnetization and specific heat down to 0.4 K was carried out on $Lu_{1-x}Zr_xB_{12}$ samples with concentration of zirconium $0.78 \leq x \leq 1$ [5, 6]. In this paper we have studied solid solutions with values of concentration $x \leq 0.45$ by precise *ac*-susceptibility measurements down to 50 mK.

## 2. Experimental details

High quality single crystalline samples of $Lu_{1-x}Zr_xB_{12}$ with concentration x = 0.00, 0.04, 0.10, 0.20 and 0.45 were prepared by vertical crucible-free inductive floating zone melting technique in an inert argon atmosphere. The exact ratio of Zr/Lu ions concentration was determined by scanning electron microscopy. To obtain the superconducting phase diagrams down to very low temperatures, a new *ac*-susceptometer for homebuilt dilution $^3$He-$^4$He minirefrigerator was designed, constructed and tested. The primary coil from NbTi superconducting wire ($T_c \approx 9$ K) was used to generate an excitation field of about 0.04 Oe by *ac*-current $I_P = 10^{-4}$ A and frequency f = 401 Hz. Usual values of detected voltage induced in the secondary coil (produced from Cu wire) were $U_i \approx 10^{-6}$ V. All samples had the same dimensions of 2 x 2 x 0.5 mm$^3$ (*a* x *b* x *c*). The external magnetic field was applied along the *c*-axis ([001] direction).

## 3. Results and discussion

For all studied samples as the first, the virgin curve i.e. temperature dependence of induced voltage $U_i$ in zero external magnetic field was measured to determine the critical temperature $T_c(0)$. Then measurements of temperature and field dependences of $U_i$ were carried out. Representative field dependences of the induced voltage signal $U_i(H)$ for $LuB_{12}$ at various stable temperatures are presented on Figure 1. The critical fields $H_c$ were defined as the midpoint of very sharp step-like superconducting phase transition of $U_i(H)$ dependence. The resulting temperature dependence of critical field $H_c(T)$ for $LuB_{12}$ is displayed in Figure 2. To our knowledge, it is the first detailed observation of the superconducting phase diagram of $LuB_{12}$ down to such low temperatures. The BCS-type of superconductivity in $LuB_{12}$ is clearly demonstrated by the perfect fit of $H_c(T)$ with BCS formula: $H_c(T) = H_c(0)(1-(T/T_c)^2)$ (see Fig. 2). From this fit the value of critical magnetic field $H_c(0) \approx 20$ Oe was determined, which is in accordance with the previous predictions and measurements down to only 0.35 K [7]. Analogously to case of $LuB_{12}$, the resulting phase diagrams of $Lu_{1-x}Zr_xB_{12}$ ($x \leq 0.45$) samples were received

(see Fig. 3) from temperature and field sweeps of $U_i$, as well as the values of critical fields were determined from BCS fits. The resulting concentration dependencies of $T_c(x)$ and $H_c(x)$ are displayed in Figure 4. The best $T_c(x)$ fit for x up to 0.45 was received by a linear function with a slope of $dT_c/dx$ = + 32 mK / at.%. However, the extrapolation of this linear fit up to the maximum value x = 1 gives a critical temperature $T_c \approx 3.6$ K which is a much lower than the actual value $T_c(ZrB_{12})$ = 6 K [7]. On the contrary, the most suitable quadratic $H_c(x)$ fit provides a critical field value of $H_c$ = 3.8 kOe which is considerably higher value than the real one $H_c(ZrB_{12})$ = 580 Oe [7]. Recently, it was shown [6] that for the concentration of zirconium above 70 % (x > 0.70) a higher value of $dT_c/dx$ = + 120÷210 mK / at.% was observed. These all indicate that a crossover in the $T_c(x)$ dependence of $Lu_{1-x}Zr_xB_{12}$ compounds appears between 45 and 70 % of Zr.

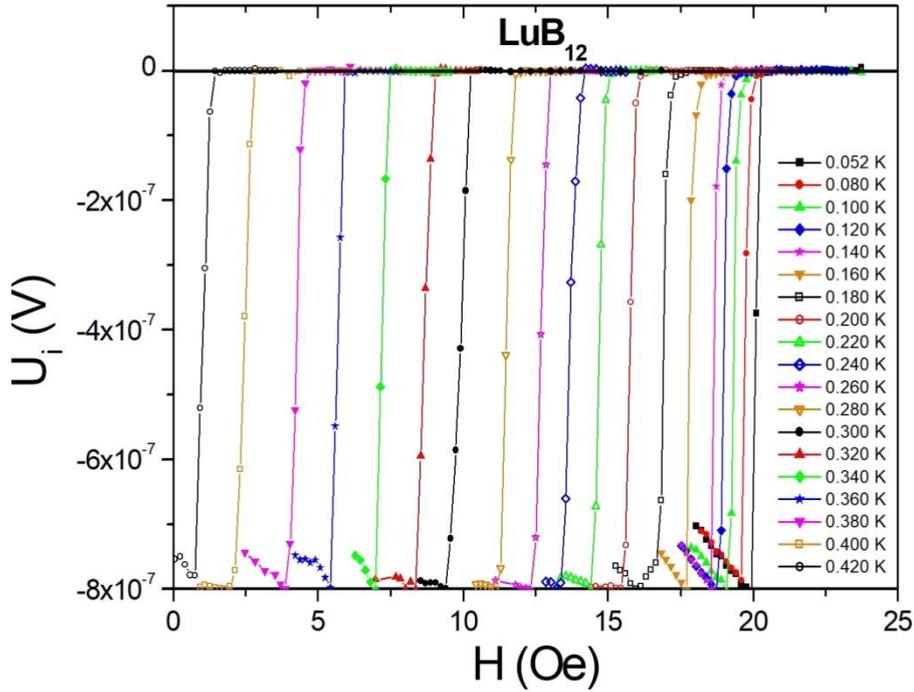

**Fig. 1.** Field dependences of induced voltage $U_i(V)$ at several constant temperatures for $LuB_{12}$ sample.

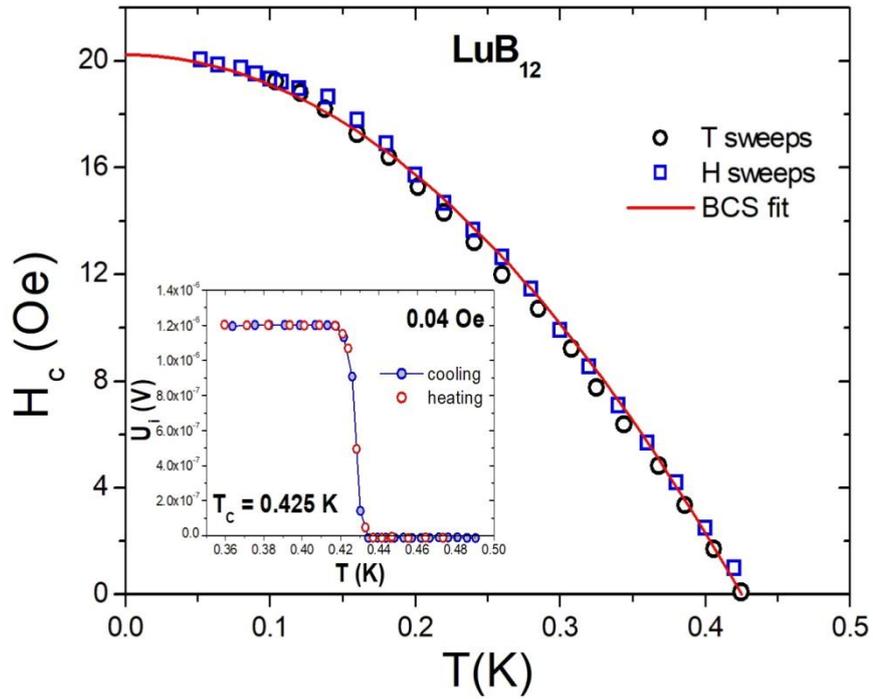

**Fig. 2.** Superconducting phase diagram $H_c(T)$ for $LuB_{12}$ received from temperature and field dependencies of ac-susceptibility, and corresponding BCS fit. Inset includes a virgin curve of induced voltage $U_i(V)$ used for the determination of exact critical temperature $T_c(0)$ values.

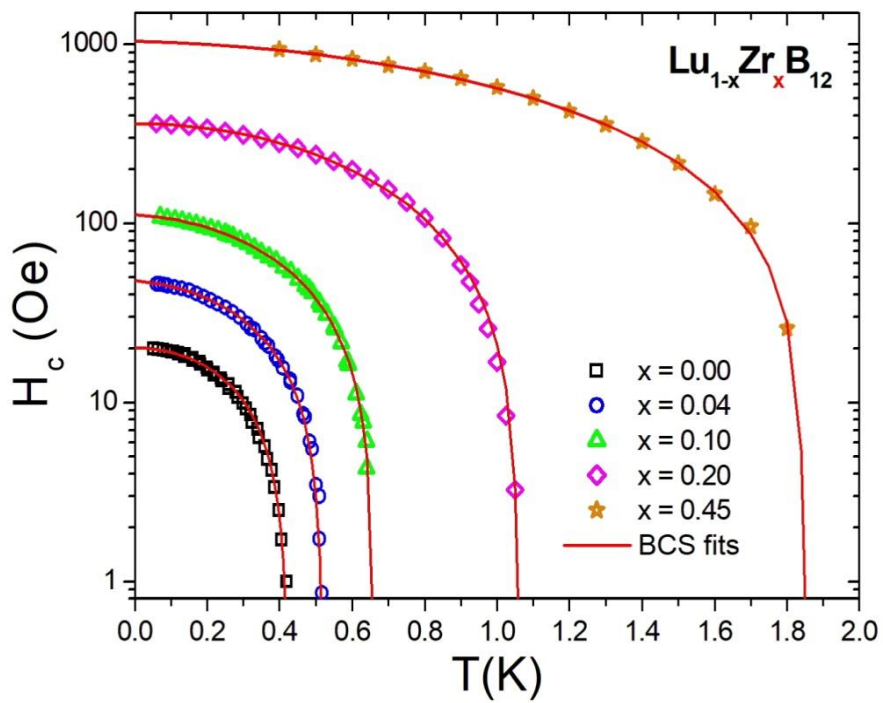

**Fig. 3.** Phase diagrams of $Lu_{1-x}Zr_xB_{12}$ solid solutions obtained from *ac*-susceptibility measurements in $^3$He-$^4$He dilution refrigerator down to 50 mK ($0 \leq x \leq 0.20$) and in $^3$He refrigerator down to 0.4 K ($x = 0.45$).

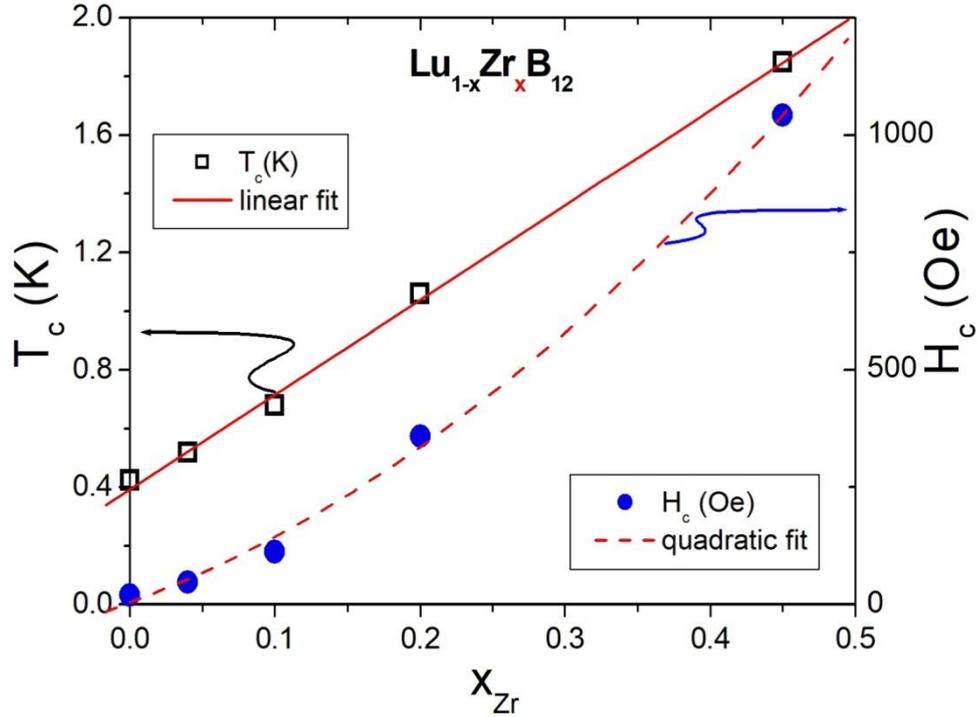

**Fig. 4.** Variation of critical temperature and critical field as a function of zirconium concentration ($0 \leq x \leq 0.45$) in $Lu_{1-x}Zr_xB_{12}$ solid solutions.

## 4. Conclusions

For the first time the superconducting phase diagram of $LuB_{12}$ as well as of $Lu_{1-x}Zr_xB_{12}$ solid solutions with $x \leq 0.45$ were studied down to very low temperatures. A linear increase of critical temperature and a quadratic increase of critical magnetic field with increasing concentration of zirconium up to 45 % were observed. To determine the exact critical concentration at which the crossover from weak to strong coupling is occurs, further precise measurements of point contact spectroscopy or specific heat have to be carried out down to very low temperatures.

**Acknowledgement**

This work was supported by the Slovak agency VEGA (grant no. 2/0032/16), EU ERDF-ITMS 26220220186 (Promatech), EU ERDF-ITMS 26110230097 (Physnet) and by the Program of Fundamental Research of the Presidium of the Russian Academy of Sciences 'Fundamental Problems of High-Temperature Superconductivity'. Liquid nitrogen for experiments was sponsored by U.S. Steel Kosice.